\documentstyle[sprocl,rotate]{article}

\input{psfig}

\newcommand{\rf}[4]{{\em {#1}} {\bf #2}, #3 (#4)}

\def\PRD{{\em Phys. Rev.} D}
\newcommand{\pr}{Phys.\ Rev.\ }
\newcommand{\prl}{Phys.\ Rev.\ Lett.\ }
\newcommand{\pl}{Phys.\ Lett.\ }

\newcommand{\np}{Nucl.\ Phys.\ }

\def\be{\begin{equation}}
\def\ee{\end{equation}}
\def\bea{\begin{eqnarray}}
\def\eea{\end{eqnarray}}
\def\Tr{{\rm Tr}\,}
\def\muhat{\hat{\mu}}
\def\qhat{\hat{q}}
\def\bra{\langle}
\def\ket{\rangle}
\newcommand{\err}[2]{\raisebox{-0.4ex}
{$\stackrel{\scriptstyle +#1}{\scriptstyle -#2}$}}

\begin{document}

\title{THE INFRARED BEHAVIOUR OF THE GLUON PROPAGATOR FROM LATTICE QCD}

\author{D.~B.~LEINWEBER, J.~I.~SKULLERUD,\footnote{UKQCD Collaboration} A. G. WILLIAMS}

\address{Department of Physics and Mathematical Physics,\\
The University of Adelaide,\\ Adelaide, SA 5005, Australia\\
http://www.physics.adelaide.edu.au/theory/}

\maketitle\abstracts{ The gluon propagator in Landau gauge is
calculated in quenched QCD on a large ($32^3\times 64$) lattice at
$\beta=6.0$.  In order to assess finite volume and finite lattice
spacing artefacts, we also calculate the propagator on a smaller
volume for two different values of the lattice spacing.  New structure
seen in the infrared region survives conservative cuts to the lattice
data, and serves to exclude a number of models that have appeared in
the literature.}

\section{Introduction}
\label{sec:intro}

The infrared behaviour of the gluon propagator is important for an
understanding of confinement.  Several studies using Dyson--Schwinger
equations~\cite{mandelstam,bp} have suggested that the gluon
propagator diverges faster than $1/q^2$ in the infrared, and it has
been claimed that this is a necessary condition for confinement.  On
the other hand, recent studies of coupled ghost and gluon
Dyson--Schwinger equations~\cite{hauck,bloch} indicate the gluon
propagator vanishes in the infrared, in agreement with older
studies~\cite{gribov,stingl} using other methods.  Other
investigations~\cite{cornwall} have also suggested a massive gluon,
yielding an infrared finite propagator.

Lattice QCD should in principle be able to resolve this issue by
first-principles, model-independent calculations.  However, lattice
studies have up to now been inconclusive,\cite{bernard,marenzoni}
since they have not been able to access sufficiently low momenta.  The
lower limit of the available momenta on the lattice is given by
$q_{\rm min} = 2\pi/L$, where $L$ is the length of the lattice.

Here we will report results using a lattice with a length of 3.3~fm in
the spatial directions and 6.7~fm in the time direction.  This gives us
access to momenta as small as 400~MeV.

\section{Lattice formalism}
\label{sec:def}

The gluon field $A_\mu$ can be extracted from the link variables
$U_\mu(x)$ using
\be
U_\mu(x) =  e^{ig_0aA_\mu(x+\muhat/2)} + {\cal O}(a^3) \, .
\ee
Inverting and Fourier transforming this, we obtain
\bea
A_\mu(\qhat) & \equiv & \sum_x e^{-i\qhat\cdot(x+\muhat/2)}
 A_\mu(x+\muhat/2) \nonumber \\
 & = & \frac{e^{-i\qhat_{\mu}a/2}}{2ig_0a}\left[\left(U_\mu(\qhat)-U^{\dagger}_\mu(-\qhat)\right)
 - \frac{1}{3}\Tr\left(U_\mu(\qhat)-U^{\dagger}_\mu(-\qhat)\right)\right] , 
\eea
where $U_\mu(\qhat)\equiv\sum_x e^{-i\qhat x}U_\mu(x)$ and
$A_\mu(\qhat)\equiv t^a A_{\mu}^a(\qhat)$.  The available momentum
values $\qhat$ are given by
\be
\qhat_\mu  = 2 \pi n_\mu/(a L_\mu), \qquad
n_\mu=0,\ldots,L_\mu-1
\ee
where $L_\mu$ is the length of the box in the $\mu$ direction.  The
gluon propagator $D^{ab}_{\mu\nu}(\qhat)$ is defined as
\be
D^{ab}_{\mu\nu}(\qhat) = \bra A^a_\mu(\qhat)
A^b_\nu(-\qhat) \ket\,/\,V \, .
\ee

\noindent In the Landau gauge, the propagator is expected to have the
form
\be
D_{\mu\nu}^{ab}(\qhat) =
\delta^{ab}(\delta_{\mu\nu}-\frac{q_{\mu}q_{\nu}}{q^2})D(q^2)
\, ,
\label{eq:landau_prop}
\ee
where the new momentum variable $q$ is defined by
\be
q_\mu \equiv \frac{2}{a}\sin\frac{\qhat_\mu a}{2} ,
\label{eq:lat-momenta}
\ee
such that at tree level, $D(q^2)$ will have the simple form
\be
D(q^2) = 1/q^2\, .
\label{eq:tree}
\ee

\section{Simulation Parameters, Finite Size Effects and Anisotropies}
\label{sec:params}

We have analysed 75 configurations at $\beta=6.0$, on a $32^3\times
64$ lattice.  Using the value of $a^{-1}=1.885$ GeV,\cite{bs}
this corresponds to a physical volume of ($3.35^3\times 6.70$)~fm.  For
comparison, we have also studied an ensemble of 125 configurations on
a smaller volume of $16^3\times 48$, with the same lattice spacing.  
An accuracy of
$\frac{1}{VN_C}\sum_{\mu,x}|\partial_{\mu}A_{\mu}|^2 <10^{-12}$ was
achieved for the Landau gauge condition 
on
both lattices.

In the following, we are particularly interested in the deviation of
the gluon propagator from the tree level form (\ref{eq:tree}).  We
will therefore factor out the tree level behaviour and plot $q^2
D(q^2)$ rather than $D(q^2)$ itself.  Furthermore, in order to
eliminate the most obvious source of lattice anisotropies, we will
plot the data as a function of the momentum variable $q$ defined in
(\ref{eq:lat-momenta}).

\begin{figure}[t]
\begin{center}
\leavevmode
\mbox{\rotate[l]{\psfig{figure=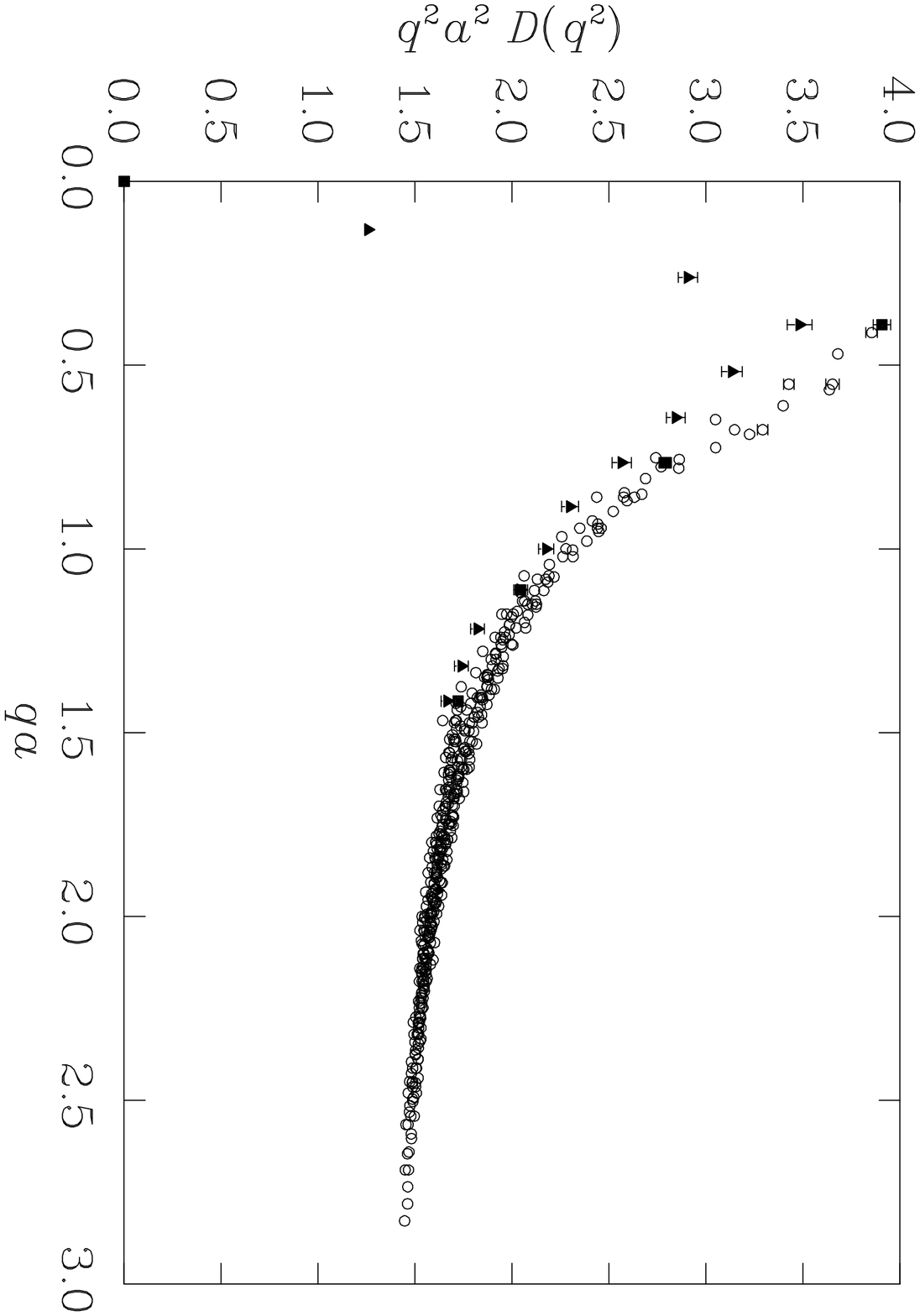,height=2.2in}}\hspace{0.5cm}
\rotate[l]{\psfig{figure=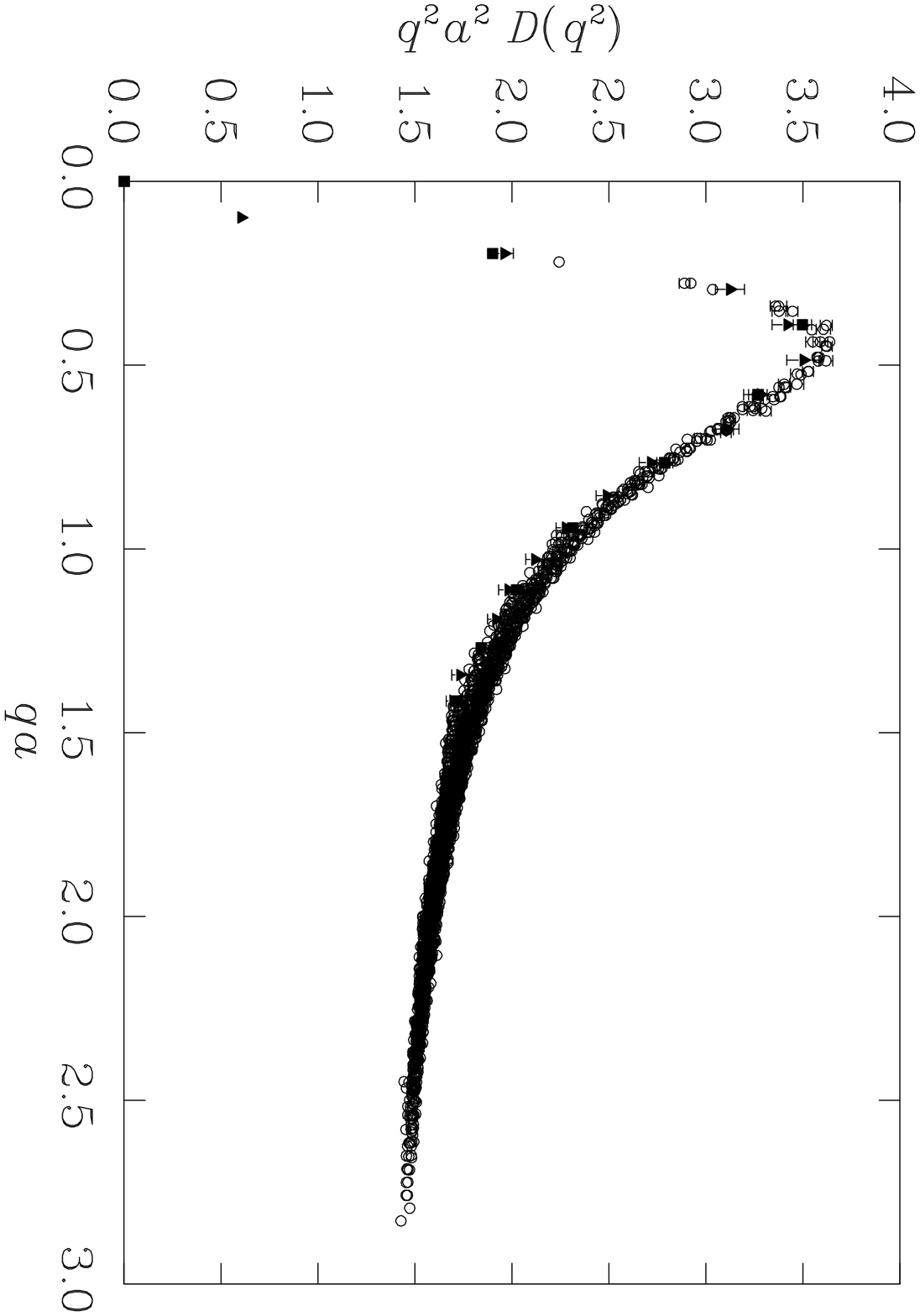,height=2.2in}}}
\end{center}
\caption{Componentwise data, for the small lattice (left) and the
large lattice (right).  The filled triangles denote momenta directed
along the time axis, while the filled squares denote momenta directed
along one of the spatial axes.}
\label{fig:cpt-data}
\end{figure}

Fig.~\ref{fig:cpt-data} shows the gluon propagator as a function of
$qa$ for both lattices, with momenta in different directions plotted
separately.  Only an averaging over the three spatial directions has
been performed.  For low momenta, there are large anisotropies in the
small lattice data, due to finite size effects.  This can be seen in
particular from the difference between points representing momenta
along the time axis and those representing momenta along the spatial
axes for $qa\sim 0.4$ and $qa\sim 0.75$.  These anisotropies are
absent from the data from the large lattice, indicating that finite
size effects here are under control.

\begin{figure}[t]
\begin{center}
\leavevmode
\rotate[l]{\psfig{figure=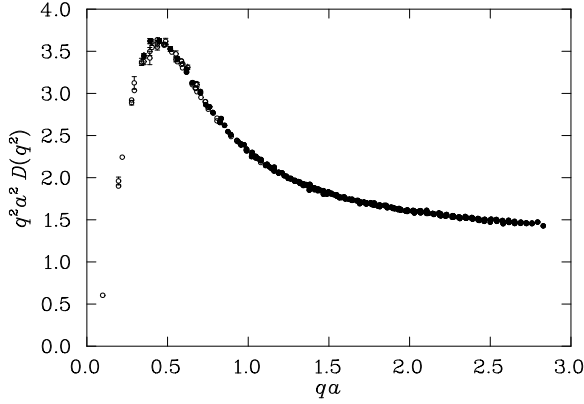,height=3.0in}}
\end{center}
\caption{The gluon propagator from the large lattice.  Only momentum
vectors with a distance from the diagonals of 2 units of momenta or
less have been kept.  The solid circles denote those points which also
lie within a cone with an opening angle of $20^\circ$.}
\label{fig:cuts}
\end{figure}

However, at higher momenta, there are anisotropies which remain for
the large lattice data, and which are of approximately the same
magnitude for the two lattices.  These anisotropies must be the result
of finite lattice spacing errors in the action, which break down the
continuum O(4) symmetry to the hypercubic symmetry.
In order to eliminate these anisotropies, we select momenta lying
within a cylinder of radius $\Delta\qhat a = 2\times 2\pi/32$ along
the 4-dimensional diagonals.  The data surviving this cut are shown in
fig.~\ref{fig:cuts}.

We also consider a further cut, keeping only those momenta which lie
within a cone of $20^\circ$ opening angle, directed along the
diagonals and with its apex at the origin.  This cut was found
necessary to remove the low-momentum anisotropies on the small
lattice.\cite{letter} The data surviving this cut are shown as filled
symbols in fig.~\ref{fig:cuts}.  It is interesting to see that even
with this conservative cut, the main qualitative features of the data
can still be observed.

\section{Model Fits}
\label{sec:models}

We have considered the following functional forms:
\bea
D(q^2) & = & \frac{Z}{(q^2)^{1+\alpha}+M^2}
\label{model:marenzoni} \\
D(q^2) & = & Z
\left[(q^2+M^2(q^2))\ln\frac{q^2+4M^2(q^2)}{\Lambda^2}\right]^{-1}
\label{model:cornwall} \\
 \mbox{where} & &
M(q^2) = M\left\{\ln\frac{q^2+4M^2}{\Lambda^2}/
\ln\frac{4M^2}{\Lambda^2}\right\}^{-6/11} \nonumber \\
D(q^2) & = & Z\left(\frac{A}{(q^2+M^2)^{1+\alpha}} + \frac{1}{q^2+M^2}\right)
\label{model3} \\
D(q^2) & = & Z\left(\frac{A}{(q^2)^{1+\alpha}+(M^2)^{1+\alpha}} +
\frac{1}{q^2+M^2}\right)
\label{model4} \\
D(q^2) & = & Z\left(A e^{-(q^2/M^2)^{\alpha}} + \frac{1}{q^2+M^2}\right)
\label{model5}
\eea

Model~(\ref{model:marenzoni}) is the one used by Marenzoni {\em et
al.}~\cite{marenzoni}  Model~(\ref{model:cornwall}) was proposed by
Cornwall.~\cite{cornwall}  Models~(\ref{model3}) and (\ref{model4})
are constructed as generalisations of (\ref{model:marenzoni}) with the
correct dimension and tree level behaviour in the ultraviolet regime.

All models were fitted to the large lattice data using the cylindrical
cut.  The lowest momentum value was excluded, as the volume dependence
of this point could not be determined.  In order to balance the
sensitivity of the fit between the high- and low-momentum region,
nearby data points within $\Delta(qa) < 0.05$ were averaged.

\begin{figure}[t]
\begin{center}
\leavevmode
\psfig{figure=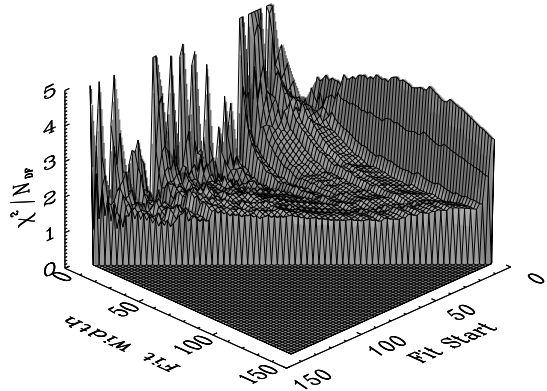,height=2.5in}
\end{center}
\vspace{-1cm}
\caption{$\chi^2$ per degree of freedom for fits to model
(\protect{\ref{model4}}). The fits are to the large lattice data with
the cylindrical cut, with nearby points averaged.  The data points are
numbered 1, \ldots, 142, with the most infrared point being number 1.
`Fit Width' denotes the number of points included in the fit.}
\label{fig:chisq-model4}
\end{figure}

Model (\ref{model4}) accounts for the data better than any of the
other models.  Fig.~\ref{fig:chisq-model4} shows $\chi^2$ per degree
of freedom for fits to this model, as a function of the starting point
of the fit (where the points are numbered 1, 2, \ldots starting from
the most infrared) and the number of points included in the fit.  The
region of interest is the right-hand section of the plots, where the
number of points in the fit is large and the infrared region is
included.  (\ref{model4}) gives a good fit (with $\chi^2/{\rm dof}
\sim 1$) for a wide range of momenta.  Only when we try to include the
most infrared points does $\chi^2/$dof go up to around 4.

\begin{figure}[p]
\begin{center}
\leavevmode
\mbox{\psfig{figure=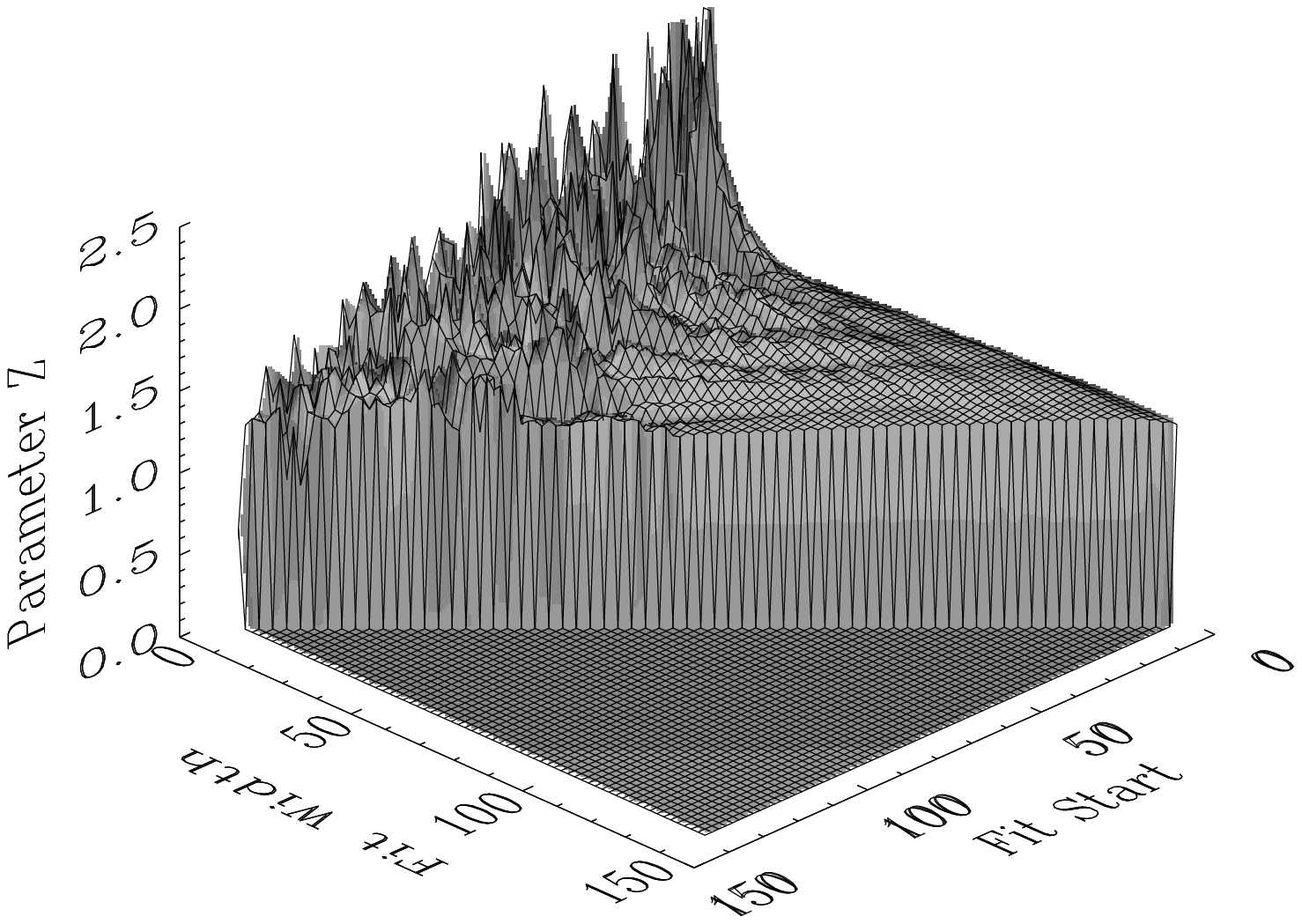,height=1.69in}\psfig{figure=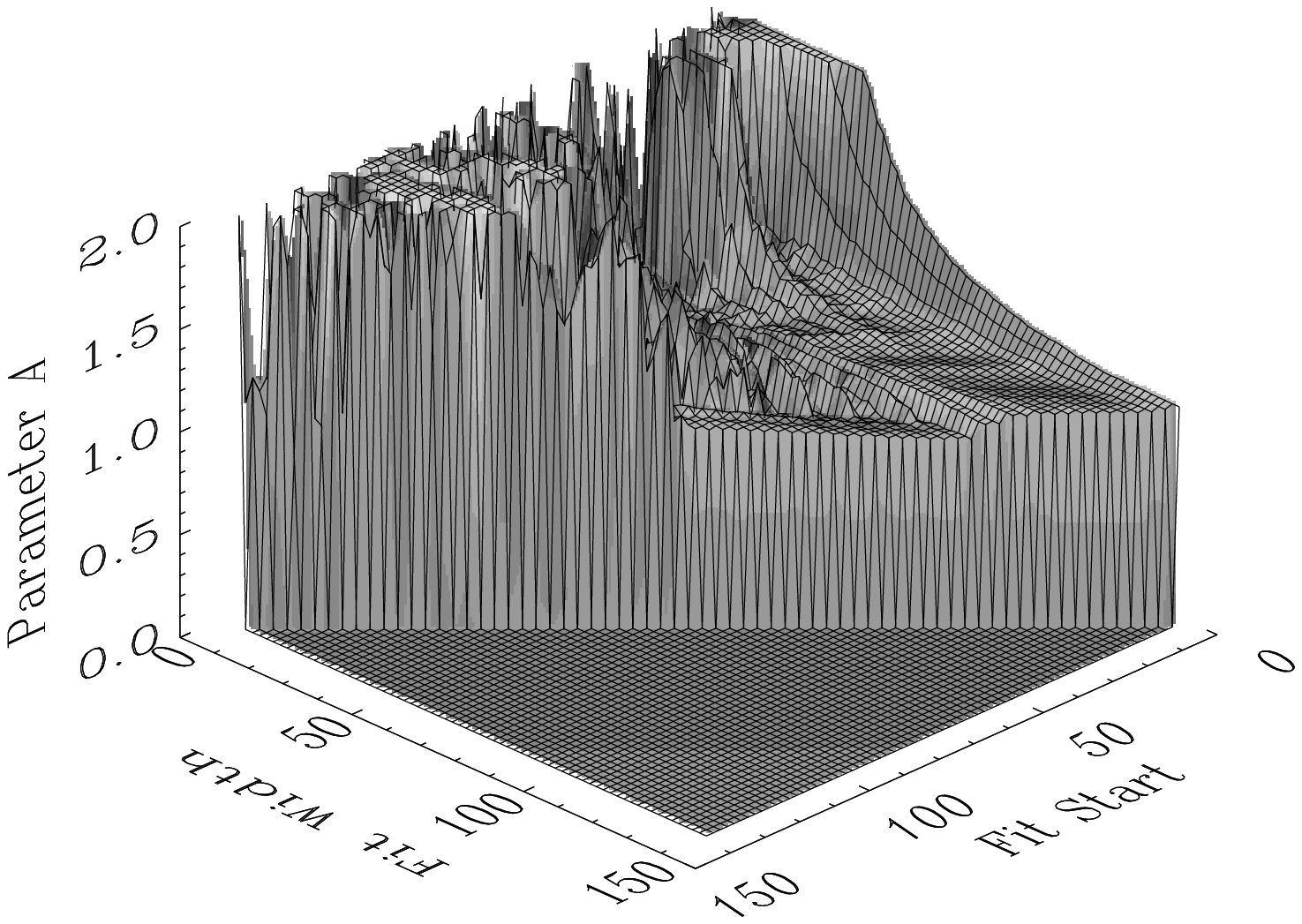,height=1.69in}}
\mbox{\psfig{figure=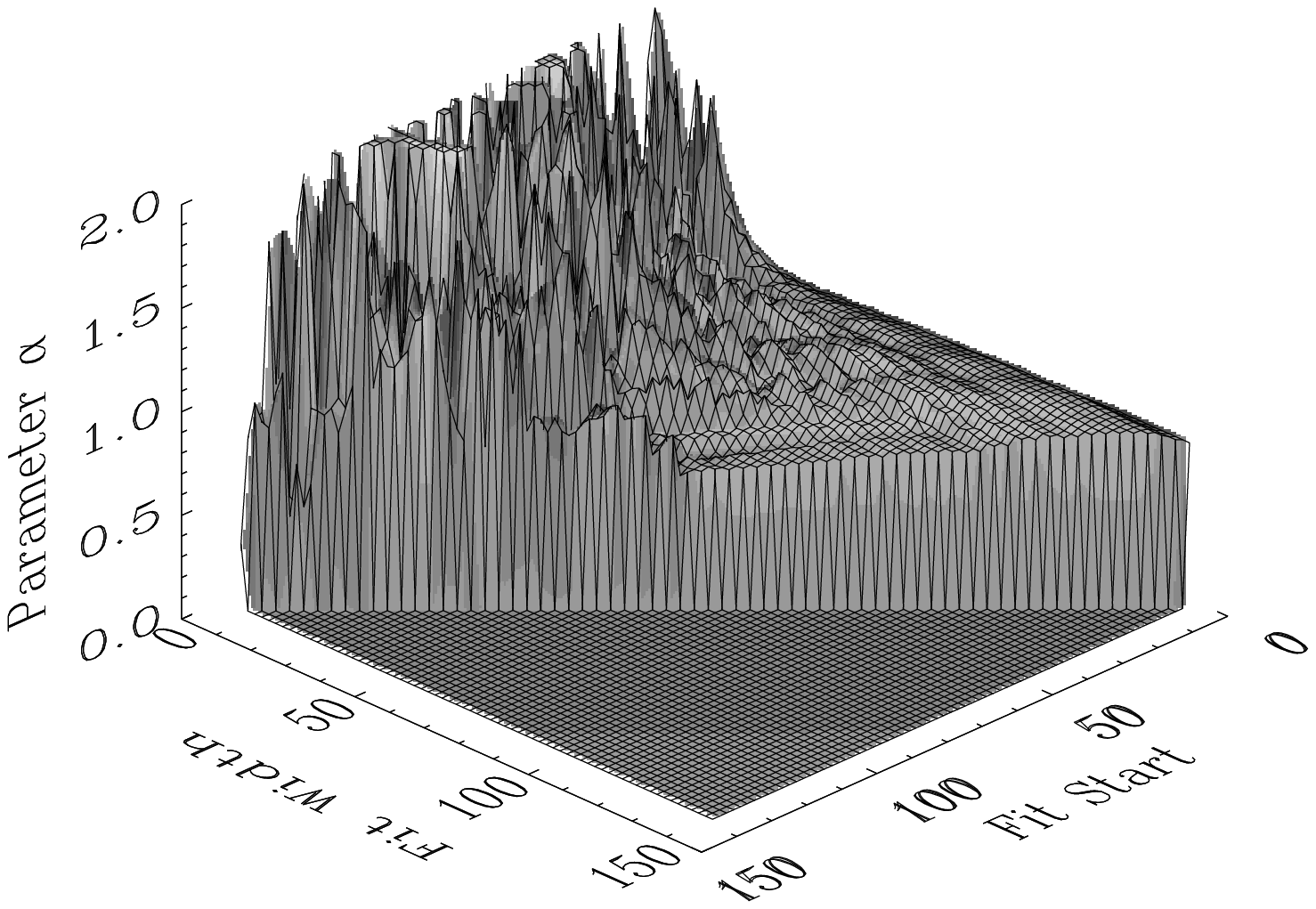,height=1.69in}\psfig{figure=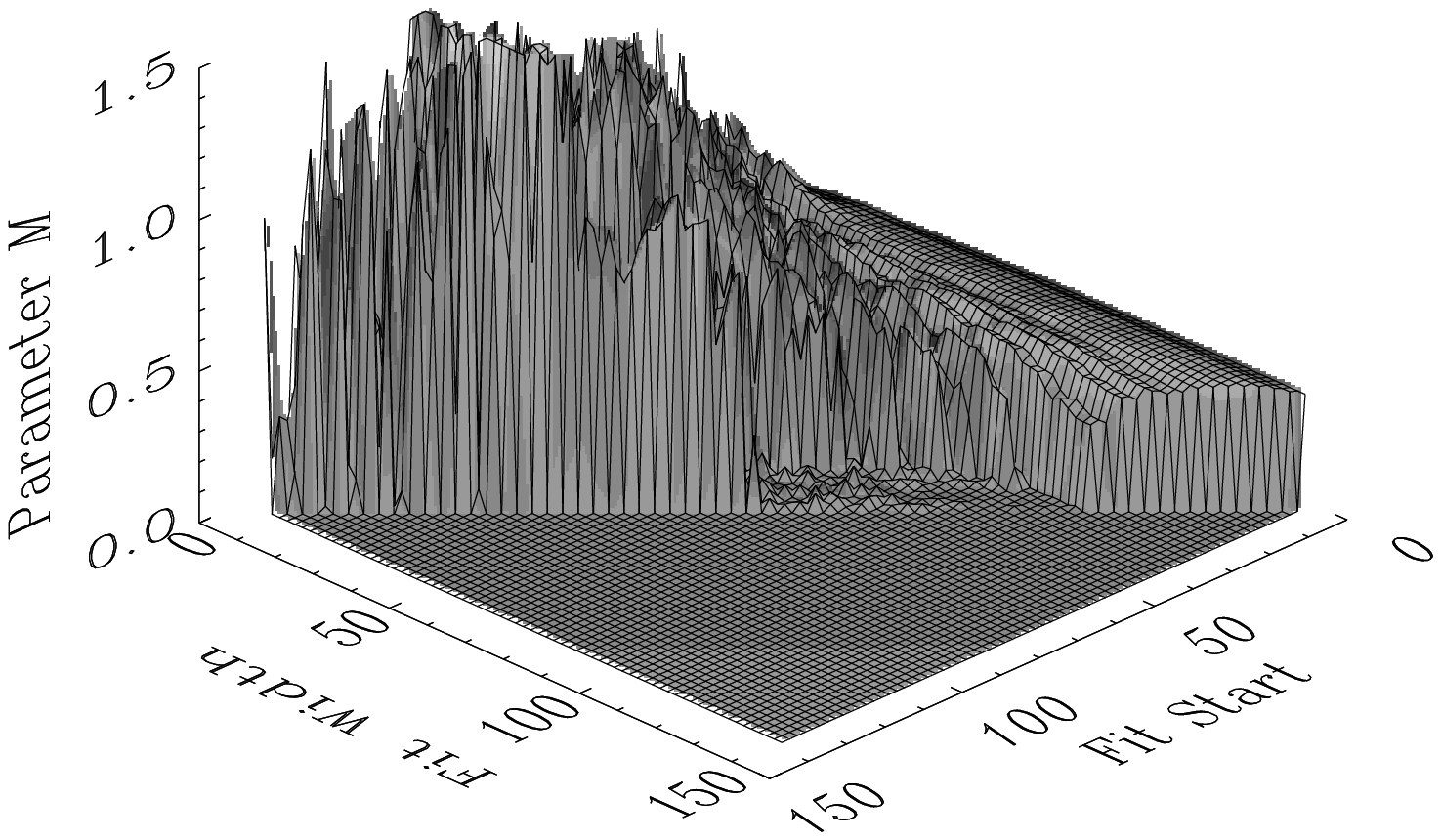,height=1.69in}}
\caption{Parameter values for fits to model (\protect{\ref{model4}}).
The axes are as in fig.~\protect{\ref{fig:chisq-model4}}.}
\end{center}
\label{fig:params-model4}
\end{figure}

The parameter values for fits to (\ref{model4}) are shown in
fig.~\ref{fig:params-model4}.  As long as a reasonable number of
points, including points in the infrared region, are included in the
fit, the values vary very little when the fitting interval is varied.
Our best estimate for the parameters is
\be
Z = 1.21\err{7}{4} 
A = 1.06\err{1}{15} 
\alpha = 0.78\err{9}{3} 
M = 0.38\err{5}{2}
\ee
where the errors denote the uncertainties in the last digit(s) of the
parameters, due to fluctuations in the parameters as the fitting
interval is varied.

\begin{figure}[p]
\mbox{\rotate[l]{\psfig{figure=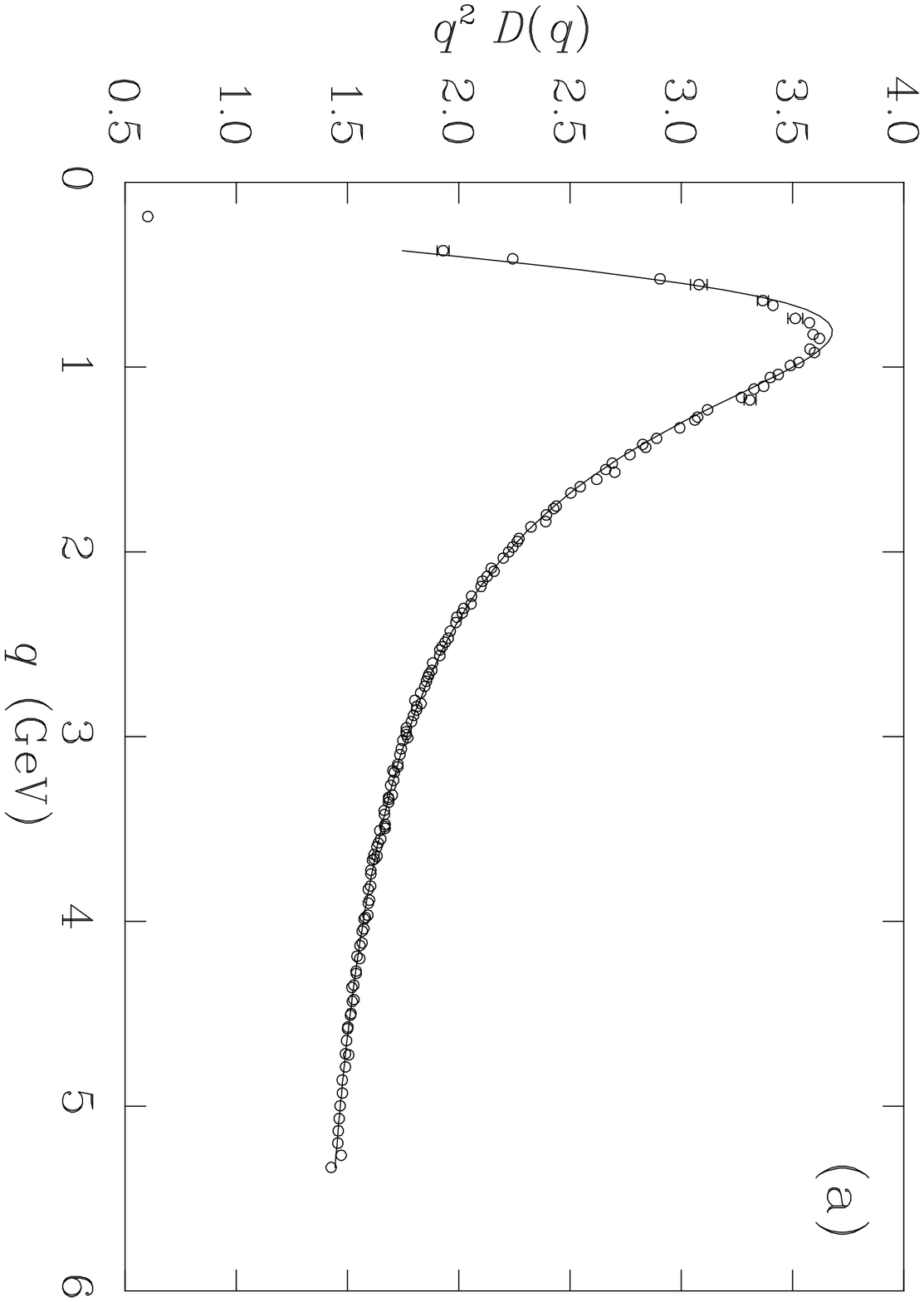,height=2.2in}}\hspace{0.5cm}
\rotate[l]{\psfig{figure=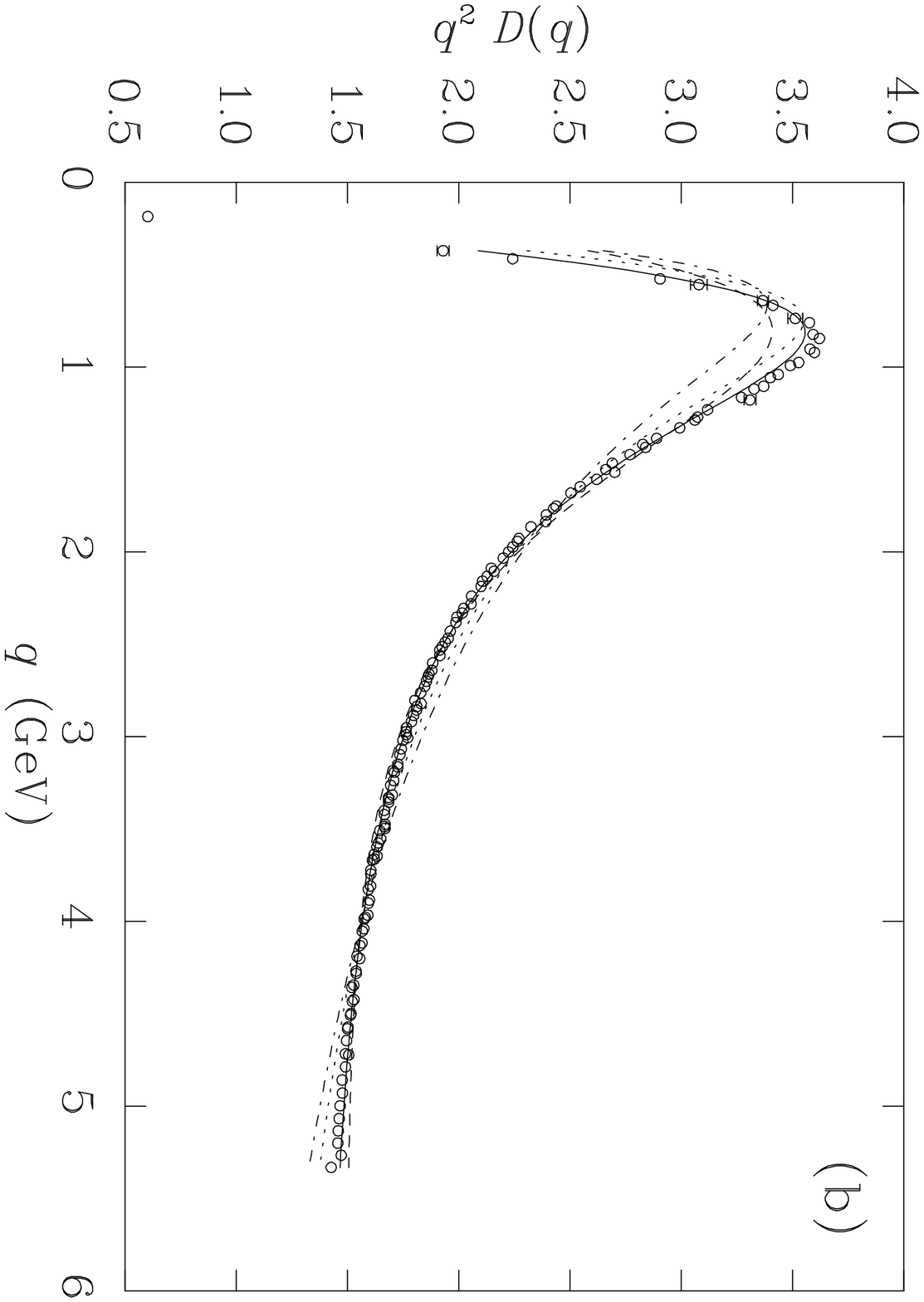,height=2.2in}}}
\caption{Fits to (a) model (\protect{\ref{model4}}); (b) models
(\protect{\ref{model:marenzoni}}) (dash--dotted line),
(\protect{\ref{model:cornwall}}) (dotted line),
(\protect{\ref{model3}}) (solid line) and (\protect{\ref{model5}})
(dashed line).}
\label{fig:fits}
\end{figure}

Fig.~\ref{fig:fits} shows the fit to (\ref{model4}) using these
parameters, as well as the best fits to the other models considered.

\section{Discussion and Outlook}
\label{sec:discuss}

We have evaluated the gluon propagator on an asymmetric lattice with a
large physical volume.  By studying the anisotropies in the data, and
comparing the data with those from a smaller lattice, we have been
able to conclude that finite size effects are under control on this
lattice.

A clear turnover in the behaviour of $q^2 D(q^2)$ has been observed at
$q \sim 1$GeV, indicating that the gluon propagator diverges less
rapidly than $1/q^2$ in the infrared, and may be infrared finite or
vanishing. 

A more detailed study, including an investigation of the tensor
structure and a detailed analysis of different functional forms, is in
progress.\cite{next}  The effect of Gribov copies is also an important
issue for consideration.
In the future, we hope to use improved actions to
perform realistic simulations at larger lattice spacings.  This would
enable us to evaluate the gluon propagator on larger physical volumes,
giving access to lower momentum values.

\section*{Acknowledgments} 

The numerical work was mainly performed on a Cray T3D at EPCC,
University of Edinburgh, using UKQCD Collaboration time under PPARC
Grant GR/K41663.  Financial support from the Australian Research
Council is gratefully acknowledged.  We thank Claudio
Parrinello for stimulating discussions.

\end{document}